\documentclass[10pt,conference]{IEEEtran}
\usepackage{epsfig,graphicx,subfigure,psfrag,amsmath,cases,bm}
\usepackage{latexsym,amssymb,algorithm,mathtools}
\usepackage{algorithmic}
\usepackage{color}
\usepackage{url}
\usepackage{scrtime}
\usepackage{stfloats}
\usepackage{tablefootnote}
\usepackage{cite}
\usepackage{amsfonts,mathabx}
\usepackage{textcomp}
\usepackage{xcolor,capt-of}
\usepackage{multirow}
\usepackage[left=0.625in,top=0.73in,bottom=0.95in,right=0.625in]{geometry}
\author{\IEEEauthorblockN {Dongfang Xu\IEEEauthorrefmark {1}, Ata Khalili\IEEEauthorrefmark {1}, Xianghao Yu\IEEEauthorrefmark {2}, Derrick Wing Kwan Ng\IEEEauthorrefmark {3},
and Robert Schober\IEEEauthorrefmark {1}\\
\IEEEauthorrefmark {1}Friedrich-Alexander-University
Erlangen-N\"urnberg, Germany;\\
\IEEEauthorrefmark {2}City University of Hong Kong, Hong Kong;
\IEEEauthorrefmark {3}The University
of New South Wales, Australia
}

\thanks{D. Xu, A. Khalili, and R. Schober are with the Institute for Digital Communications, Friedrich-Alexander-University Erlangen-N\"urnberg (FAU), Germany (email:\{dongfang.xu, ata.khalili, robert.schober\}@fau.de).
X. Yu is with Department of Electronic and Computer Engineering, the Hong Kong University of Science and Technology, Hong Kong (e-mail: eexyu@ust.hk).
D. W. K. Ng is with the School of Electrical Engineering and Telecommunications, the University of New South Wales, Australia (email: w.k.ng@unsw.edu.au).
}
}

\newtheorem{T-Prob}{Transformed Problem}

\DeclareMathOperator{\mino}{minimize}

\DeclareMathOperator{\subto}{subject\hspace*{2mm}to}

\newtheorem{Remark}{Remark}

\allowdisplaybreaks

\title{Integrated Sensing and Communication in Distributed Antenna
Networks}

\begin{document}
\maketitle

\begin{abstract}
In this paper, we investigate the resource allocation design for integrated sensing and communication (ISAC) in distributed antenna networks (DANs). In particular, coordinated by a central processor (CP), a set of remote radio heads (RRHs) provide communication services to multiple users and sense several target locations within an ISAC frame. To avoid severe interference between the information transmission and the radar echo, we propose to divide the ISAC frame into a communication phase and a sensing phase. During the communication phase, the data signal is generated at the CP and then conveyed to the RRHs via fronthaul links. As for the sensing phase, based on pre-determined RRH-target pairings, each RRH senses a dedicated target location with a synthesized highly-directional beam and then transfers the samples of the received echo to the CP via its fronthaul link for further processing of the sensing information. Taking into account the limited fronthaul capacity and the quality-of-service requirements of both communication and sensing, we jointly optimize the durations of the two phases, the information beamforming, and the covariance matrix of the sensing signal for minimization of the total energy consumption over a given finite time horizon. To solve the formulated non-convex design problem, we develop a low-complexity alternating optimization algorithm which converges to a suboptimal solution. Simulation results show that the proposed scheme achieves significant energy savings compared to the two baseline schemes. Moreover, our results reveal that for efficient ISAC in wireless networks, energy-focused short-duration pulses are favorable for sensing while low-power long-duration signals are preferable for communication.

\end{abstract}
\section{Introduction}
Recently, integrated sensing and communication (ISAC) has emerged as a promising technique to not only provide high data rate services to conventional communication users, but also to support various environment-aware Internet-of-Things (IoT) applications \cite{9705498}. In particular, enabled by dual-functional radar-communication (DFRC) base stations (BSs), ISAC empowers radar systems and wireless communication systems to share the scarce spectrum and expensive hardware. Motivated by these advantages, various works in the literature have proposed advanced ISAC techniques to improve the performance of wireless communication systems and provide on-demand sensing services. For instance, the authors of \cite{8288677} developed an efficient beamforming policy for synthesizing desired radar beam pattern in a multiuser ISAC system. Also, in \cite{9838753}, the authors investigated the joint design of information and sensing beamforming to maximize ISAC performance in a single-user communication system in the presence of multiple targets. However, due to the severe signal attenuation experienced by radar echoes, the potential of ISAC cannot be fully unleashed by the conventional co-located-antenna networks considered in \cite{8288677,9838753}. In practice, the sensing quality and the maximum sensing range of radar systems are determined by the power level of the received radar echo \cite{skolnik2008radar}. Yet, due to the round-trip attenuation of the echo signal, ISAC systems employing co-located antennas have to allocate a significant portion of their limited power supply to compensate for the severe path loss. For instance, a sensing target located at a distance of $200$ meters from the transmitter causes a free-space round-trip path loss of roughly $170$ dB for a $2.4$ GHz sensing signal. Assuming an ISAC system with $5$ MHz bandwidth, to ensure the echo can be distinguished from the noise having a typical power spectral density of $-174$ dBm/Hz, the sensing signal transmit power has to be on the order of several kilowatt \cite{skolnik2008radar}. This results in an unaffordably high energy requirement, especially for the conventional cellular BSs considered in \cite{8288677,9838753}, for which the power budget is usually less than $100$ watts. This creates a bottleneck for achieving high-quality ISAC.
\par
A promising approach to overcome this challenge are distributed antenna networks (DANs). Specifically, a DAN comprises a central processor (CP) and a group of low-cost remote radio heads (RRHs), where the RRHs are distributed across the network and exchange data with the CP through dedicated fronthaul links \cite{5706317}. In fact, this distributed network architecture allows shortening the distance between the DFRC transmitters and the sensing targets, which is beneficial for reducing the required transmit power and extending the maximum sensing range. Therefore, compared to conventional co-located-antenna networks, DANs are more suitable for the realization of practical ISAC systems. Yet, in DANs, the information data for the communication users and the received echo of the target required for sensing information extraction have to be exchanged via fronthaul links between the RRHs and the CP. In practice, the capacity of the fronthaul links is limited due to the finite bandwidth available \cite{7106496}. Hence, the capacity limitation of the fronthaul links has to be carefully considered for system design to fully unleash the potential of DAN-based ISAC. On the other hand, most existing works on ISAC design, e.g., \cite{8288677,9838753}, propose to transmit the information signal and the sensing signal simultaneously via a mono-static radar-based BS. Yet, \cite{8288677,9838753} only focus on joint information beamforming and sensing signal design and do not consider the radar echo reception, which may not ensure reliable echo signal detection for practical ISAC systems. In particular, in \cite{8288677,9838753}, the desired radar echo arrives at the BS before the information transmission ends, which causes severe self-interference (SI) at the BS. However, the SI cancellation techniques designed for conventional FD communication systems may not be able effectively suppress the SI to below the target echo power since the echo signal is severely attenuated by the round-trip path loss. To overcome this difficulty, in this paper, we propose to perform sensing and communication in orthogonal time slots and adopt pulse radar to be able to flexibly adjust the sensing range. In particular, according to pulse radar theory, the sensing range depends on the durations of the sensing pulse and the received echo \cite{skolnik2008radar}. As a result, the system designer has to carefully divide the available sensing time into two parts to ensure reliable echo detection at the transmitter. Although employing a DAN architecture to further enhance the performance of ISAC systems seems promising, the corresponding resource allocation design needed to facilitate high-quality sensing and communication with limited fronthaul capacity remains an open problem.

\par
Motivated by the above observations, in this paper, we propose to achieve efficient ISAC by deploying a DAN and investigate the corresponding resource allocation algorithm design. To avoid interference between information transmission and the radar echo, we adopt an ISAC frame structure comprising two phases, i.e., a communication phase and a sensing phase. Specifically, we jointly design the durations of both phases, the beamforming vectors, and the covariance matrix of the sensing signal for minimization of the total energy consumption within an ISAC frame. The corresponding resource allocation algorithm design is formulated as a non-convex optimization problem taking into account the limited fronthaul capacity and the quality-of-service (QoS) requirements of both communication and sensing. Furthermore, we propose a low-complexity alternating optimization (AO)-based iterative algorithm which is guaranteed to converge to a high-quality solution of the considered optimization problem.
\par
\textit{Notation:} Vectors and matrices are denoted by boldface lower case and boldface capital letters, respectively. $\mathbb{R}^{N\times M}$ and $\mathbb{C}^{N\times M}$ denote the space of $N\times M$ real-valued and complex-valued matrices, respectively. $|\cdot|$ and $||\cdot||_2$ denote the absolute value and the $l_2$-norm of their arguments, respectively. $||\cdot||_0$ is the $l_0$-norm of a vector counting the number of non-zero entries in the vector; $(\cdot)^T$ and $(\cdot)^H$ stand for the transpose and the conjugate transpose of their arguments, respectively. $\mathbf{I}_{N}$ refers to the identity matrix of dimension $N$. $\mathbb{H}^{N}$ denotes the set of complex Hermitian matrices of dimension $N$. $\mathrm{Tr}(\cdot)$ and $\mathrm{Rank}(\cdot)$ refer to the trace and rank of their arguments, respectively. $\mathbf{A}\succeq\mathbf{0}$ indicates that $\mathbf{A}$ is a positive semidefinite matrix. $\mathrm{diag}(\mathbf{a})$ represents a diagonal matrix whose main diagonal elements are given by vector $\mathbf{a}$; $\mathcal{CN}(0 ,\sigma^2)$ specifies the distribution of a circularly symmetric complex Gaussian (CSCG) random variable with mean $0$ and variance $\sigma^2$. $\overset{\Delta }{=}$ and $\sim$ stand for ``defined as'' and ``distributed as'', respectively. $\mathcal{E}\left \{ \cdot \right \}$ denotes statistical expectation.
\section{System Model}
In this section, we first introduce the proposed distributed antenna ISAC system and then present the corresponding signal and fronthaul models.
\begin{figure}[t] 
\centering\includegraphics[width=3.4in]{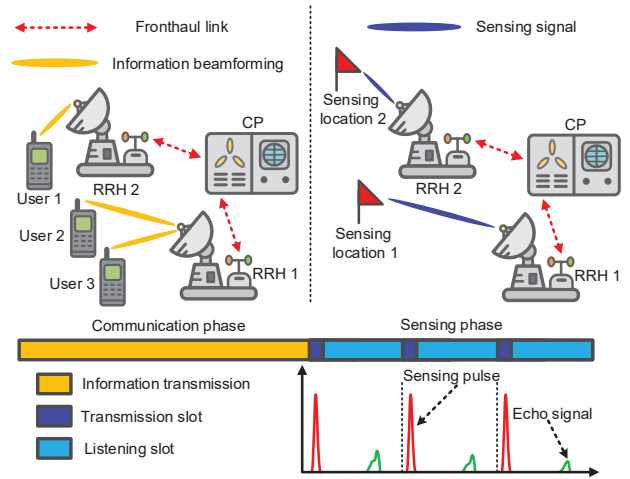}
\caption{Illustration of a DAN-based multiuser ISAC system comprising a CP, $J=2$ RRHs, $K=3$ users, and $L=2$ desired sensing locations. The CP and the RRHs exchange the information data and the samples of the received echo via fronthaul links. The color bar indicates the proposed frame structure comprising a communication phase and a sensing phase. In the communication phase, the RRHs jointly implement beamforming to serve all the users in a collaborative manner. Also, the sensing phase is divided into $M=3$ rounds, where each round contains a transmission slot for pulse emission and a listening slot for echo reception. The red and green signals in the bottom-right figure represent the sensing pulse and echo signal, respectively.}\label{fig:1 system}
\end{figure}
\subsection{DAN-Based ISAC System Model}
We consider a DAN-based ISAC system which comprises a CP, $J$ RRHs, and $K$ users which are single-antenna devices, cf. Fig. 1. Each RRH is equipped with a uniform linear array with $N_{\mathrm{T}}$ antenna elements and is connected with the CP via an individual fronthaul link. Also, we assume that there are $L$ desired sensing locations in the ISAC system, at each of which one potential target may exist. Moreover, as the core unit of the network, the CP is expected to perform all computations needed for operating the ISAC system. We note that when employing simultaneous information signal and sensing signal transmission, the desired radar echo usually arrives at the BS before the information transmission ends. In this case, the ISAC transmit signal may cause severe SI to the received echo signal at the BS, which is challenging to mitigate with conventional SI cancellation techniques. To overcome this difficulty, we propose to divide each ISAC frame into two phases, i.e., a communication phase and a sensing phase, having durations $(1-\eta)T$ and $\eta T$, respectively, where $0<\eta<1$ denotes the time allocation variable. In particular, in the communication phase, the CP computes the beamforming vectors for the users, and forwards the information data of the users and the control signals for resource allocation to the RRHs via the fronthaul links.\footnote{In practice, the fronthaul links can be established via out-of-band radio frequency links or free-space optical links \cite{7106496}.} On the other hand, we propose to assign a dedicated RRH to each sensing location to facilitate high-quality sensing based on an energy-focused beam. Assuming $J\geq L$, the RRH assignment policy is pre-designed in an offline manner\footnote{In practice, the RRH assignment policy can be based on the system geometry and the sensing application scenario, e.g., by pairing RRHs and sensing locations based on distance.} and is utilized for the online resource allocation optimization, which facilitates a low-complexity ISAC design. Hence, there are $L$ pre-determined RRH-target pairs in the considered system. During the sensing phase, the CP computes the covariance matrix of the sensing signal for each pre-designed RRH-target pair and forwards the resulting resource allocation policy to each RRH via a fronthaul link. Then, all RRH concurrently emit highly-directional beams towards their respective paired target locations in the transmission slot of duration of $t_p$ before switching to the listening mode to receive the echoes. To facilitate high-quality sensing, the sensing phase is divided into $M$ sensing slots \cite{skolnik2008radar}, each of which has a duration of $\frac{\eta T}{M}$. The received echoes are transferred to the CP via the fronthaul link to facilitate the acquisition of fine-resolution sensing information. Besides, to investigate the performance upper bound of the considered system, we assume perfect channel state information of the users and the desired target locations is available at the CP. For notational simplicity, we collect the indices of the RRHs, users, and desired target locations in sets $\mathcal{J}=\left \{1,\cdots, J \right \}$, $\mathcal{K}=\left \{1,\cdots ,K \right \}$, and $\mathcal{L}=\left \{1,\cdots ,L \right \}$, respectively.
\begin{Remark}
We note that the proposed RRH-target pairing policy facilitates power-efficient high-quality sensing in the considered DAN-based ISAC system. In particular, this policy simplifies the sensing beam design, i.e., it facilitates the use of highly-directional beam patterns, which helps avoid severe clutter and interference between different RRH-target pairs \cite{skolnik2008radar}. Also, the proposed policy allows us to sense each target location via the nearest RRH, which potentially improves the network power efficiency. As an alternative strategy, all the $J$ RRHs may concurrently sense all target locations. However, this would result in a complicated multi-beam pattern synthesis problem and may lead to severe crosstalk between different RRHs, which would degrade sensing quality. 
\end{Remark}
\subsection{Signal Model}
In the communication phase, $K$ independent data streams are transmitted simultaneously to the $K$ users. In particular, a beamforming vector dedicated to user $k$, $\mathbf{w}_{j,k}\in \mathbb{C}^{N_{\mathrm{T}}\times 1}$, is generated at the CP and is conveyed to RRH $j$ via fronthaul link $j$. For ease of presentation, we define a super-vector $\mathbf{w}_{k}=\big[(\mathbf{w}_{1,k})^T,(\mathbf{w}_{2,k})^T,\cdots,(\mathbf{w}_{J,k})^T\big]^T\in \mathbb{C}^{N_{\mathrm{T}}J\times 1}$ to collect all the beamforming vectors for serving user $k$. The transmit signal vector at the RRHs is given by $\underset{k\in\mathcal{K}}{\sum}\mathbf{w}_kb_k$, where $b_k\in\mathbb{C}$ denotes the information-bearing symbol intended for user $k$ and we assume $\mathcal{E}\{\left |b_k \right|^2\}=1$, $\forall\mathit{k} \in \mathcal{K}$, without loss of generality. The received signal at user $k$ is given by
\begin{equation}
y_{\mathrm{U}_k}=\mathbf{h}_k^H\mathbf{w}_kb_k+\mathbf{h}_k^H\underset{r\in\mathcal{K}\setminus  \left\{k\right\}}{\sum}\mathbf{w}_rb_r+n_{\mathrm{U}_k}.
\end{equation}
Here, vector $\mathbf{h}_k\in \mathbb{C}^{N_{\mathrm{T}}J\times 1}$ represents the channel between the $J$ RRHs and user $k$. $n_{\mathrm{U}_k}\sim\mathcal{CN}(0,\sigma_{\mathrm{U}_k}^2)$ denotes the equivalent additive white Gaussian noise (AWGN) at user $k$ with variance $\sigma_{\mathrm{U}_k}^2$. 
\par
In the sensing phase, for the $l$-th RRH-target pair, a sensing signal $\mathbf{s}_l\in \mathbb{C}^{N_{\mathrm{T}}\times 1}$ is transmitted by the RRH to illuminate the desired target location. In this paper, we assume sensing signal $\mathbf{s}_l$ is transmitted in each of the $M$ sensing rounds and model $\mathbf{s}_l$ as a CSCG random vector with zero mean and covariance matrix $\mathbf{S}_l\succeq \mathbf{0}$ \cite{9838753}. We note that $\mathbf{s}_l$ is generated locally at the associated RRH based on the optimized resource allocation policy, i.e., the fronthaul links are not involved. The received signal at the $l$-th RRH is modeled as \cite{8579200}
\begin{equation}
\label{sig_model}
\mathbf{y}_l=\underbrace{\mathbf{g}_{l,l}\mathbf{g}_{l,l}^H\mathbf{s}_l}_{\text{Desired echo}}+\underbrace{\underset{q\in\mathcal{L}\setminus  \left\{l\right\}}{\sum}\big(\underset{p\in\mathcal{L}}{\sum}\mathbf{g}_{l,p}\mathbf{g}_{q,p}^H\mathbf{s}_q+\mathbf{g}_{l,q}\mathbf{g}_{l,q}^H\mathbf{s}_l\big)}_{\text{Clutter}}+\mathbf{n}_l,
\end{equation}
where $\mathbf{n}_l\sim\mathcal{CN}(\mathbf{0},\sigma_l^2\mathbf{I}_{N_{\mathrm{T}}})$ denotes the AWGN at RRH $l$ with noise variance $\sigma_l^2$. Moreover, similar to \cite{8288677}, in this paper, we assume pure line-of-sight (LoS) channels between the desired sensing locations and the RRHs. In particular, the channel between RRH $l$ and location $q$ is denoted by vector $\mathbf{g}_{l,q}\in\mathbb{C}^{N_{\mathrm{T}}\times 1}$, $\forall l, q\in\mathcal{L}$, and is given by
\begin{equation}
\mathbf{g}_{l,q}=\sqrt{\beta_{l,q}}\mathbf{a}(\phi_{l,q}),
\end{equation}
where $\beta_{l,q}\in\mathbb{R}$ and $\mathbf{a}(\phi_{l,q})\in\mathbb{C}^{N_{\mathrm{T}}\times 1}$ denote the channel gain and the steering vector between RRH $l$ and location $q$, respectively. In particular, $\beta_{l,q}$ is given by $\beta_{l,q}=\gamma_q\frac{\varrho_0 }{d^2_{l,q}}$, where $d_{l,q}$ is the distance between RRH $l$ and location $q$ and $\gamma_q\in\mathbb{R}$ is the radar cross-section of target $q$ \cite{8579200}. The value of constant $\varrho_0=(\frac{c}{4\pi f_c})^2$ depends on the system center frequency $f_c$ and the speed of light $c$. Moreover, vector $\mathbf{a}(\phi_{l,q})$ is given by $\mathbf{a}(\phi_{l,q})=\Big[1,e^{j2\pi\omega\mathrm{sin}\phi_{l,q}},\cdots,e^{j2\pi\omega(N_\mathrm{T}-1)\mathrm{sin}\phi_{l,q}}\Big]^T$ with $\phi_{l,q}$ and $\omega$ being the angle of departure (AoD) from RRH $l$ to location $q$ and the normalized spacing between adjacent antenna elements \cite{8288677}, respectively. To facilitate the presentation, we define matrix $\mathbf{G}_{l,p,q}\in\mathbb{C}^{N_{\mathrm{T}}\times N_{\mathrm{T}}}$ to characterize the channel of the two-hop link from RRH $q$ to RRH $l$ via target location $p$, where $\mathbf{G}_{l,p,q}=\mathbf{g}_{l,p}\mathbf{g}_{q,p}^H$, $\forall l,p,q\in\mathcal{L}$ \cite{8579200}.
\subsection{Fronthaul Model}
Via fronthaul links, the information data of all users and samples of the received echoes are exchanged between the RRHs and the CP in the communication and sensing phases, respectively. Specifically, in the communication phase, we model the required fronthaul capacity\footnote{In this paper, we assume that for each fronthaul link, a fixed amount of fronthaul capacity has already been reserved for forwarding the resulting resource allocation policy from the CP to the RRHs. Thus, this part is not considered in \eqref{com_front}.} for RRH $j$ as follows \cite{7106496}
\begin{equation}
\label{com_front}
 C_j^{\mathrm{C}}=\frac{\underset{k\in\mathcal{K}}{\sum }\left\|\left\| \mathbf{w}_{j,k}\right\|_2 \right\|_0N^{\mathrm{C}}_k}{(1-\eta)TW_\mathrm{F}}\hspace*{2mm}\mathrm{(bits/s/Hz)},\hspace*{1mm}j\in\mathcal{J}.
\end{equation}
Here, constants $N^{\mathrm{C}}_k$ and $W_\mathrm{F}$ denote the number of bits dedicated to user $k$ during one ISAC frame and the bandwidth of the fronthaul links, respectively. To facilitate the presentation, we define $\overline{R}_{\mathrm{req}_k}=\frac{N^{\mathrm{C}}_k}{TW_\mathrm{F}}$ (bits/s/Hz) as the required data rate for delivering the data intended for user $k$ via a fronthaul link to an RRH. Moreover, the value of $\underset{k\in\mathcal{K}}{\sum }\left\|\left\| \mathbf{w}_{j,k}\right\|_2\right\|_0$ belongs to set $\mathcal{K}$ and represents the number of users served by RRH $j$. In practice, due to the limited capacity of the fronthaul links, each RRH may not be able to simultaneously serve all $K$ users. Instead, based on the resource allocation policy, the CP can facilitate partial cooperation, i.e., delivering the information data dedicated to user $k$ only to a subset of the RRHs. This can be achieved by setting $\mathbf{w}_{i,k}=\mathbf{0}$, $i\in\mathcal{J}$, which excludes RRH $i$ from participating in the joint information transmission to user $k$. On the other hand, for target sensing, the RRH emits beams of duration $t_p$ and then switches to the listening mode to receive the corresponding echoes until the end of the sensing round. According to pulse radar theory \cite{skolnik2008radar}, the minimum sensing range and the maximum sensing range of the considered ISAC system are given by
\begin{eqnarray}
R_{\mathrm{min}}=\frac{ct_p}{2}\hspace*{2mm}\mbox{and}\hspace*{2mm} R_{\mathrm{max}}=\frac{c(\frac{\eta T}{M}-t_p)}{2}\label{max_dist},
\end{eqnarray}
respectively. In particular, $R_{\mathrm{min}}$ and $R_{\mathrm{max}}$ denote the minimum and maximum distances that a radar with pulse duration $t_p$ and subsequent reception duration $(\frac{\eta T}{M}-t_p)$ can detect and still produce reliable information, respectively. We note that unlike in the communication phase, where the RRHs receive only data symbols from the CP, in the sensing phase, the RRH has to first sample and quantize the received echo based on the desired sensing resolution, and then forward the quantized data to the CP \cite{9737357}. Based on the above considerations, for each sensing round of duration $\frac{\eta T}{M}-t_p$, we model the required fronthaul capacity for conveying the sampled and quantized echoes from a RRH to the CP as follows \cite{lu2021performance}
\begin{equation}
\label{sen_front}
    C^{\mathrm{S}}=\frac{(R_{\mathrm{max}}-R_{\mathrm{min}})N_{\mathrm{b}}}{\Delta R (\frac{\eta T}{M}-t_p) W_\mathrm{F}}\hspace*{2mm}\mathrm{(bits/s/Hz)}.
\end{equation}
Here, parameter $\Delta R>0$ denotes the pre-defined target resolution of the radar in meter and measures the ability of a radar to distinguish between the desired target and other objects in the vicinity.\footnote{In practice, the value of $\Delta R$ is determined by the width of the pulse, type of target, and efficiency of the radar \cite{skolnik2008radar}.} Besides, constant $N_{\mathrm{b}}$ denotes the number of bits required for quantizing the echo without causing saturation.

\section{Problem Formulation}
In this section, we formulate the resource allocation algorithm design as a non-convex optimization problem, after defining the adopted QoS metrics for the considered ISAC system.
\subsection{Performance Metrics}
During the communication phase, the achievable rate (bits/s/Hz) of user $k$ is given by 
\begin{eqnarray}
R_k=\mathrm{log}_2\Big(1+\frac{\left|\mathbf{h}_k^H\mathbf{w}_k\right|^2}{\underset{r\in\mathcal{K}\setminus  \left\{k\right\}}{\sum}\left|\mathbf{h}_k^H\mathbf{w}_r\right|^2+\sigma_{\mathrm{U}_k}^2}\Big).
\end{eqnarray}
\par
On the other hand, to facilitate high-quality sensing during the sensing phase, the desired sensing location has to be illuminated by an energy-focusing beam with low side lobe leakage such that the desired echoes can be easily distinguished from clutter. To this end, we discretize the angular domain $[0,\hspace*{1mm}2\pi]$ into $I$ directions and generate the ideal beam pattern $\{P(\theta_i)\}_{i=1}^{I}$ offline, where $P(\theta_i)$ denotes the beam pattern power in direction $\theta_i$. Specifically, for the $l$-th RRH-target pair, $\{P_l(\theta_i)\}_{i=1}^{I}$ is given by
\begin{equation}
P_l(\theta_i)=\left\{\begin{matrix}
1, & \hspace*{6mm}\left|\theta_i -\phi_{l,l}\right| \leq \frac{\psi_l}{2} \\ 
0, & \mbox{otherwise}\\ 
\end{matrix}\right., \hspace*{1mm}\forall l\in\mathcal{L},\label{idealbeam}
\end{equation}
where $\psi_l$ is the desired beamwidth of the ideal beam pattern for the $l$-th RRH-target pair. Since the ideal beam pattern is difficult to generate in practice, we approximate it by suitably choosing the covariance matrix of the sensing signal \cite{9737357}, i.e., $\mathbf{S}_l$. To quantify the beam pattern matching accuracy, in this paper, we adopt the difference between the ideal beam pattern and the actual beam pattern as a performance metric for sensing as follows
\begin{equation}
D_l\big(\mathbf{S}_l,\xi_l\big)=\sum_{i=1}^{I}\left| P_l(\theta_i)-\xi_l\mathbf{a}^H(\theta_i)\mathbf{S}_l\mathbf{a}(\theta_i) \right|, \forall l\in\mathcal{L},
\end{equation}
where $\xi_l$ is a scaling factor for the $l$-th RRH-target pair.\footnote{Different from some existing works, e.g., \cite{9838753}, in this paper, we use $\xi_l$ to scale the actual beam pattern instead of the ideal beam pattern. By doing so, we can flexibly control the beam pattern mismatch error to satisfy different beam synthesis accuracy requirements for a given ideal beam pattern \cite{skolnik2008radar}.} Moreover, to facilitate reliable detection of the desired echo signal at the RRH, we consider two additional performance metrics for sensing. On the one hand, to ensure the desired echo signal can be effectively captured by the paired RRH, we consider the received echo power at RRH $l$ which is given by $\mathrm{Tr}(\mathbf{G}_{l,l,l}\mathbf{S}_l\mathbf{G}_{l,l,l}^H)$. On the other hand, to effectively suppress the clutter, we consider the interference at RRH $l$, i.e., $I_l$, which can be calculated based on \eqref{sig_model} as follows\footnote{In this paper, we consider the received echo power and interference at RRHs to explicitly show the effect of the round-trip path loss on the sensing signal. They can also be combined as signal-to-interference-plus-noise ratio \cite{skolnik2008radar}.}
\begin{eqnarray}
    I_l&\hspace*{-2mm}=\hspace*{-2mm}&\underset{q\in\mathcal{L}\setminus  \left\{l\right\}}{\sum}\Big(\underset{p\in\mathcal{L},p'\in\mathcal{L}}{\sum}\mathrm{Tr}\big(\mathbf{G}_{l,p,q}\mathbf{S}_q\mathbf{G}_{l,p',q}^H\big)\notag\\
    &\hspace*{-2mm}+\hspace*{-2mm}&\underset{q'\in\mathcal{L}}{\sum}\mathrm{Tr}\big(\mathbf{G}_{l,q,l}\mathbf{S}_l\mathbf{G}_{l,q',l}^H\big)+\mathrm{Tr}\big(\mathbf{G}_{l,l,l}\mathbf{S}_l\mathbf{G}_{l,q,l}^H\big)\Big).
\end{eqnarray}
\subsection{Optimization Problem Formulation}
In this paper, our objective is to minimize the total energy consumption of the considered system over a given time horizon of duration $T$ and for a pre-designed RRH-target assignment policy. In particular, given  the ideal beam pattern gain $\{P_l(\theta_i)\}_{i=1}^{M}$, then the desired time allocation variable $\eta$, the sensing signal duration $t_p$, the beamforming vectors $\mathbf{w}_k$, and the covariance matrix of sensing signal $\mathbf{S}_l$ can be obtained by solving the following problem 
\begin{eqnarray}
\label{prob1}
&&\hspace*{-6mm}\underset{\substack{\mathbf{S}_l\in\mathbb{H}^{N_{\mathrm{T}}},\mathbf{w}_{k},\\ \eta, t_p, \xi_l>0}}{\mino} \,\, \,\, \hspace*{4mm}f\overset{\Delta}{=}(1-\eta)T\underset{k\in\mathcal{K}}{\sum}\left\| \mathbf{w}_k \right\|_2^2+Mt_p\underset{l\in\mathcal{L}}{\sum}\mathrm{Tr}(\mathbf{S}_l)\notag\\
&&\hspace*{-6mm}\subto\hspace*{5mm}
\mbox{C1:}\hspace*{1mm}R_k\geq \frac{\overline{R}_{\mathrm{req}_k}}{(1-\eta)},\hspace*{1mm}\forall k,\notag\\
&&\hspace*{16mm}\mbox{C2:}\hspace*{1mm}\mathrm{Tr}(\mathbf{G}_{l,l,l}\mathbf{S}_l\mathbf{G}_{l,l,l}^H)\geq P_{\mathrm{req}_l},\hspace*{1mm}\forall l,\notag\\
&&\hspace*{16mm}\mbox{C3:}\hspace*{1mm}I_l\leq P_{\mathrm{tol}_l},\hspace*{1mm}\forall l,\notag\\
&&\hspace*{16mm}\mbox{C4:}\hspace*{1mm}D_l\big(\mathbf{S}_l,\xi_l\big)\leq \varepsilon_l,\hspace*{1mm}\forall l,
\notag\\
&&\hspace*{16mm}\mbox{C5:}\hspace*{1mm}\underset{k\in\mathcal{K}}{\sum }\left\| \mathbf{w}_{j,k}\right\|_2^2\leq P^{\mathrm{max}}_{\mathrm{T}_j},\hspace*{1mm}\forall j,\notag\\
&&\hspace*{16mm}\mbox{C6:}\hspace*{1mm}\mathrm{Tr}( \mathbf{S}_l)\leq P^{\mathrm{max}}_{\mathrm{T}_l},\hspace*{1mm}\forall l,
\notag\\
&&\hspace*{16mm}\mbox{C7:}\hspace*{1mm}\mathbf{S}_l\succeq\mathbf{0},\hspace*{1mm}\forall l,\notag\\
&&\hspace*{16mm}\mbox{C8:}\hspace*{1mm}\frac{\underset{k\in\mathcal{K}}{\sum }\left\|\left\| \mathbf{w}_{j,k}\right\|_2 \right\|_0\overline{R}_{\mathrm{req}_k}}{1-\eta}\leq C^{\mathrm{max}}_{\mathrm{F}_j},\hspace*{1mm}\forall j,
\notag\\
&&\hspace*{16mm}\mbox{C9:}\hspace*{1mm}\frac{(R_{\mathrm{max}}-R_{\mathrm{min}})N_{\mathrm{b}}}{\Delta R (\frac{\eta T}{M}-t_p) W_\mathrm{F}}\leq C^{\mathrm{max}}_{\mathrm{F}_l},\hspace*{1mm}\forall l,\notag\\
&&\hspace*{16mm}\mbox{C10:}\hspace*{1mm}t_{\mathrm{min}}\leq t_p,
\notag\\
&&\hspace*{16mm}\mbox{C11:}\hspace*{1mm}\frac{c(\frac{\eta T}{M}-t_p)}{2}\geq d_{l,l}\geq \frac{ct_p}{2},\notag\\
&&\hspace*{16mm}\mbox{C12:}\hspace*{1mm}0< \eta < 1.
\end{eqnarray}
In constraint C1, we impose a lower bound on the achievable rate of user $k$, i.e., $\overline{R}_{\mathrm{req}_k}$, to guarantee satisfactory communication services within an ISAC frame. To ensure that RRH $l$ can effectively detect the echo of the associated target, we restrict the minimum echo power strength and the maximum tolerable interference to be above $P_{\mathrm{req}_l}$ and below $P_{\mathrm{tol}_l}$, respectively, as specified in constraints C2 and C3, respectively. To generate favorable highly-directional beams, in constraint C4, we restrict the difference between the ideal beam pattern and the actual beam pattern to be below an error tolerance $\epsilon_l$  \cite{xu2022robust}. Constant $P^{\mathrm{max}}_{\mathrm{T}_j}$ in C5 denotes the per RRH power budget for all $J$ RRHs during the communication phase. Similarly, during the sensing phase, we limit the maximum transmit power of the RRHs to $P^{\mathrm{max}}_{\mathrm{T}_l}$, $\forall l \in \mathcal{L}$, as shown in constraint C6.  $\mathbf{S}_l\in\mathbb{H}^{N_{\mathrm{T}}}$ and constraint C7 guarantee that $\mathbf{S}_l$ is a covariance matrix. In constraint C8, we constrain the capacity consumption of fronthaul link $j$ to the maximum capacity allowance $C^{\mathrm{max}}_{\mathrm{F}_j}$. Constraint C9 is imposed to ensure that for the $l$-th RRH-target pair, the delivery of the echo via fronthaul link $l$ does not violate the maximum capacity allowance $C^{\mathrm{max}}_{\mathrm{F}_l}$ in each sensing round. Also, constraint C10 specifies the minimum duration of the sensing pulse, i.e., $t_{\mathrm{min}}$. In practice, due to the limitations of the hardware, a minimum time interval is required to switch between transmission and listening \cite{skolnik2008radar}. For each RRH-target pair, the desired target location should be included in the radar sensing range \cite{skolnik2008radar}, as indicated in constraint C11. Constraint C12 indicates the range of the time allocation variable. 
\par
We note that optimization problem \eqref{prob1} is non-convex. Specifically, the non-convexity stems from the coupled optimization variables in the objective function, the fractional functions in constraint C1, and the non-convex $l_0$-norm in constraint C8. In general, it is challenging to find the globally optimal solution to problem \eqref{prob1} in polynomial time. To overcome this difficulty, in the next section, we develop an AO-based algorithm which can produce a high-quality solution of \eqref{prob1} with low computational complexity.
\section{Solution of the Optimization Problem}
In this section, we tackle optimization problem \eqref{prob1}. In particular, we first recast \eqref{prob1} into an equivalent form and then, develop a computationally-efficient AO algorithm for handling the resulting problem.
\subsection{Problem Reformulation}
We start by defining the beamforming matrices $\mathbf{W}_k=\mathbf{w}_k\mathbf{w}_k^H$, $\forall k\in\mathcal{K}$. Moreover, we introduce an auxiliary binary optimization variable $\zeta_{j,k}$ and recast \eqref{prob1} in equivalent form as follows
\begin{eqnarray}
\label{prob2}
&&\hspace*{-10mm}\underset{\substack{\mathbf{S}_l\in\mathbb{H}^{N_{\mathrm{T}}},\mathbf{W}_{k}\in\mathbb{H}^{N_{\mathrm{T}}J},\\ \eta, t_p,\xi_l>0,\zeta_{j,k}}}{\mino} \,\, \,\, \hspace*{2mm}f\notag\\
&&\hspace*{-5mm}\subto\hspace*{6mm}\mbox{C1-C7},\mbox{C9-C12},\notag\\
&&\hspace*{18mm}\mbox{C8:}\hspace*{1mm}\underset{k\in\mathcal{K}}{\sum }\zeta_{j,k}\overline{R}_{\mathrm{req}_k}\leq (1-\eta)C^{\mathrm{max}}_{\mathrm{F}_j},\hspace*{1mm}\forall j,
\notag\\
&&\hspace*{18mm}
\mbox{C13:}\hspace*{1mm}\mathrm{Tr}(\mathbf{W}_{k}\mathbf{D}_{j})\leq \zeta_{j,k}P^{\mathrm{max}}_{\mathrm{T}_j},\notag\\
&&\hspace*{18mm}\mbox{C14:}\hspace*{1mm}\zeta_{j,k}\in\left\{0,1 \right\},\hspace*{1mm}\forall j,\hspace*{1mm}\forall k,\notag\\
&&\hspace*{18mm}\mbox{C15:}\hspace*{1mm}\mathbf{W}_k\succeq \mathbf{0},\hspace*{1mm}\forall k,\notag\\
&&\hspace*{18mm}\mbox{C16:}\hspace*{1mm}\mathrm{Rank}(\mathbf{W}_k) \leq 1,\hspace*{1mm}\forall k.
\end{eqnarray}
Here, to facilitate the presentation, we define a block diagonal matrix $\mathbf{D}_j\in\mathbb{R}^{N_{\mathrm{T}}J\times N_{\mathrm{T}}J}$ which is given by $\mathbf{D}_j\overset{\Delta }{=}\mathrm{diag}(\underbrace{0,\cdots,0}_{(j-1)N_{\mathrm{T}}},\underbrace{1,\cdots,1}_{N_{\mathrm{T}}},\underbrace{0,\cdots,0}_{(J-j)N_{\mathrm{T}}})$. Moreover, constraints C13 and C14 are two auxiliary constraints to ensure the equivalence between \eqref{prob1} and \eqref{prob2}. The auxiliary binary optimization variable $\zeta_{j,k}$ can be regarded as a binary indicator for whether the data of user $k$ is conveyed to RRH $j$ for information transmission or not. In particular, for $\mathrm{Tr}(\mathbf{W}_{k}\mathbf{D}_{j})>0$, the data of user $k$ occupies $\frac{\overline{R}_{\mathrm{req}_k}}{1-\eta}$ bit/s/Hz of the capacity of fronthaul link $j$, leading to $\zeta_{j,k}=1$ and vice versa. On the other hand, $\mathbf{W}_k\in\mathbb{H}^{N_{\mathrm{T}}J}$ and constraints C15 and C16 are imposed to guarantee that $\mathbf{W}_k=\mathbf{w}_k\mathbf{w}_k^H$ holds after optimization. As \eqref{prob1} and \eqref{prob2} are equivalent, next, we focus on the optimization problem in \eqref{prob2} and develop a corresponding resource allocation algorithm.
\begin{algorithm}[t]
\caption{AO-Based Algorithm}
\begin{algorithmic}[1]
\small
\STATE Set iteration index $i=1$, convergence tolerance $0<\delta\ll1$, and penalty factor $\mu\gg1$, initialize the optimization variables $\mathbf{W}_k^{(i)}$, $\mathbf{S}_l^{(i)}$, $\eta^{(i)}$, $t_p^{(i)}$, $\xi_l^{(i)}$, and $\zeta_{j,k}^{(i)}$.
\REPEAT
\STATE Solve the relaxed version of \eqref{prob3} for given $\eta^{(i)}$, $t_p^{(i)}$, $\xi_l^{(i)}$, $\zeta_{j,k}^{(i)}$, and update the solution $\mathbf{W}_k^{(i+1)}$, $\mathbf{S}_l^{(i+1)}$, $\zeta_{j,k}^{(i+1)}$
\STATE Solve \eqref{prob4} for $\mathbf{W}_k=\mathbf{W}_k^{(i+1)}$, $\mathbf{S}_l=\mathbf{S}_l^{(i+1)}$, $\zeta_{j,k}=\zeta_{j,k}^{(i+1)}$ and obtain $\eta^{(i+1)}$, $t_p^{(i+1)}$, $\xi_l^{(i+1)}$, $\zeta_{j,k}^{(i+1)}$
\STATE Set $i=i+1$
\UNTIL $\frac{\left|\overline{f}^{(i)}-\overline{f}^{(i-1)}\right|}{\overline{f}^{(i)}}\leq \delta$
\end{algorithmic}
\end{algorithm}
\subsection{Handling Binary Constraint C14}
The binary constraint C14 is an obstacle to efficiently tackling \eqref{prob2}. To circumvent this obstacle, we replace constraint C14 with the following two equivalent constraints
\begin{equation}
    \mbox{C14a:}\hspace*{1mm}0\leq \zeta_{j,k}\leq 1,\hspace*{2mm}\mbox{C14b:}\hspace*{1mm}\underset{j\in\mathcal{J}}{\sum }\underset{k\in\mathcal{K}}{\sum }(\zeta_{j,k}-\zeta_{j,k}^2)\leq 0.
\end{equation}
Here, constraint $\mbox{C14b}$ is the difference of two convex functions, which makes the constraint non-convex. To overcome this difficulty, we resort to a penalty-based method \cite{nocedal1999numerical} and rewrite the objective function in \eqref{prob1} as
\begin{equation}
    \overline{f}=f+\mu\underset{j\in\mathcal{J}}{\sum }\underset{k\in\mathcal{K}}{\sum }(\zeta_{j,k}-\zeta_{j,k}^2), \label{obj2}
\end{equation}
where $\mu\gg 1$ is the penalty factor  penalizing the violation of constraint $\mbox{C14b}$. Then, we apply successive convex approximation to overcome the non-convexity of constraint $\mbox{C14b}$. In particular, we construct a global underestimator for the term $\zeta_{j,k}^2$ by expanding it to a first-order Taylor series at feasible point $\zeta_{j,k}^{(i)}$ and rewrite $\overline{f}$ in \eqref{obj2} as follows
\begin{equation}
    \overline{f}=f+\mu\underset{j\in\mathcal{J}}{\sum }\underset{k\in\mathcal{K}}{\sum }\big[\zeta_{j,k}- 2\zeta_{j,k}^{(i)}\zeta_{j,k}+(\zeta_{j,k}^{(i)})^2\big],
\end{equation}
where superscript $i$ indicates the iteration index of the optimization algorithm.
\vspace*{-2mm}
\subsection{Alternating Optimization-Based Algorithm}
Next, for handling the coupled optimization variables in \eqref{prob2}, we divide the optimization variables into two blocks, i.e., $\left\{ \mathbf{S}_l,\mathbf{W}_k,\zeta_{j,k} \right\}$ and $\left\{ \eta,t_p,\xi_l,\zeta_{j,k} \right\}$, and solve the resulting two subproblems in an alternating manner.
\par
\textit{Block 1}: We first obtain $\left\{ \mathbf{S}_l,\mathbf{W}_k,\zeta_{j,k} \right\}$ by solving the following problem
\begin{eqnarray}
\label{prob3}
&&\hspace*{-6mm}\underset{\substack{\mathbf{S}_l\in\mathbb{H}^{N_{\mathrm{T}}},\zeta_{j,k},\\\mathbf{W}_{k}\in\mathbb{H}^{N_{\mathrm{T}}J}}}{\mino} \,\, \,\, \hspace*{2mm}\overline{f}\notag\\
&&\hspace*{-6mm}\subto\hspace*{2mm}\mbox{C1:}\hspace*{1mm}\chi_k\big(\underset{r\in\mathcal{K}\setminus  \left\{k\right\}}{\sum}\mathrm{Tr}(\mathbf{H}_k\mathbf{W}_r)+\sigma_{\mathrm{U}_k}^2\big)\notag\\
&&\hspace*{18mm}\leq \mathrm{Tr}(\mathbf{H}_k\mathbf{W}_k),\hspace*{1mm}\forall k,
\notag\\
&&\hspace*{13mm}\mbox{C2-C8},\mbox{C13},\mbox{C14a},\mbox{C15},\mbox{C16},
\end{eqnarray}
where scalar $\chi_k$ is defined as $\chi_k=2^{\frac{\overline{R}_{\mathrm{req}_k}}{1-\eta}}-1$ for notational simplicity. By employing semidefinite relaxation (SDR), we omit the only non-convex constraint in \eqref{prob3}, i.e., constraint C16, and solve the remaining problem by applying standard convex problem solvers such as CVX. The tightness of the SDR is revealed in the following theorem.
\par
\textit{Theorem 1:}\hspace*{1mm}The optimal beamforming matrix $\mathbf{W}^*_k$ of the rank constraint-relaxed version of \eqref{prob3} is a unit-rank matrix.
\par
\textit{Proof:} The proof of Theorem 1 follows similar steps as the proof in \cite[Appendix A]{yu2020irs}, and is thus omitted here due to page limitation.
\par
\textit{Block 2}: For given $\left\{ \mathbf{S}_l,\mathbf{W}_k,\zeta_{j,k} \right\}$, we obtain $\left\{ \eta,t_p,\xi_l,\zeta_{j,k} \right\}$ by solving the following convex optimization problem
\begin{eqnarray}
\label{prob4}
&&\hspace*{-6mm}\underset{\eta,t_p,\xi_l>0,\zeta_{j,k}}{\mino} \,\, \,\, \hspace*{2mm}\overline{f}\notag\\
&&\hspace*{-6mm}\subto\hspace*{2mm}\hspace*{2mm}\mbox{C1},\mbox{C4},\mbox{C8-C13},\mbox{C14a}.
\end{eqnarray}
\par
The proposed algorithm is summarized in \textbf{Algorithm 1}. Note that the values of $\overline{f}$ in \eqref{prob3} and \eqref{prob4} are monotonically non-increasing in each iteration of \textbf{Algorithm 1}. Moreover, according to \cite{bezdek2002some}, for a sufficiently large $\mu$, \textbf{Algorithm 1} is guaranteed to converge to a stationary value of the objective function of \eqref{prob1} in polynomial time, producing a high-quality solution for \eqref{prob1}. The computational complexity of \textbf{Algorithm 1} is given by $\mathcal{O}\Big(\mathrm{log}(\frac{1}{\delta})\big((KJ+2K+J)N^3_{\mathrm{T}}J^{3}+(KJ+2K+J)^2N^2_{\mathrm{T}}J^2+4LN_{\mathrm{T}}^3+(4L)^2N_{\mathrm{T}}^2\big)\Big)$, where $\mathcal{O}\left ( \cdot  \right )$ is the big-O notation and $\delta$ is the convergence tolerance of \textbf{Algorithm 1}.
\section{Simulation Results}
\begin{figure}[t]
\centering\includegraphics[width=2.0in]{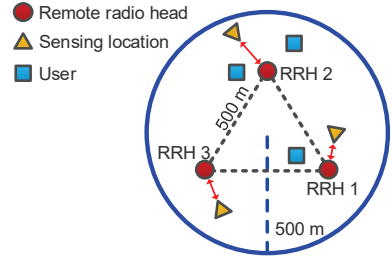}
\caption{Simulation setup for a distributed antenna multiuser ISAC system comprising $J=3$ RRHs, $K=3$ users, and $L=3$ sensing locations. The red double arrows indicate the pre-determined RRH-target pairs. }\label{fig:system_simulation} 
\end{figure}
\begin{table}[t]\caption{System simulation parameters.}\label{ISAC_parameters}\footnotesize
\newcommand{\tabincell}[2]{\begin{tabular}{@{}#1@{}}#2\end{tabular}}
\centering
\renewcommand{\arraystretch}{1.5}
\begin{tabular}{|l|l|l|}
\hline
    \hspace*{-1mm}$\sigma_{\mathrm{U}_k}^2$& Noise power at user $k$& $-104$ dBm \\
\hline
    \hspace*{-1mm}$T$& ISAC frame length & $1$ ms \\
\hline
    \hspace*{-1mm}$P^{\mathrm{max}}_{\mathrm{T}_j}$& Maximum transmit power at each RRH & $46.5$ dBm \cite{access2020radio}\\
\hline
    \hspace*{-1mm}$\overline{R}_{\mathrm{req}_k}$&  Required achievable rate of users & $2$ bits/s/Hz \\
\hline
    \hspace*{-1mm}$P_{\mathrm{req}_l}$& Minimum required echo power & $-90$ dBm \\
\hline
    \hspace*{-1mm}$P_{\mathrm{tol}_l}$& Maximum interference tolerance & $-100$ dBm \\
\hline
    \hspace*{-1mm}$C^{\mathrm{max}}_{\mathrm{F}}$& Maximum fronthaul link capacity & $5.5$ bits/s/Hz \cite{7106496} \\
\hline
    \hspace*{-1mm}$N_b$&  Number of bits for quantizing echoes & $4$ \\
\hline
    \hspace*{-1mm}$M$&  Number of rounds in sensing phase & $100$ \\
\hline
    \hspace*{-1mm}$\Delta R$ &  Sensing resolution  & $10$ m \cite{skolnik2008radar} \\
\hline
    \hspace*{-1mm}$W_{\mathrm{F}}$ & Fronthaul link bandwidth & $10$ MHz \cite{huq2016backhauling} \\
\hline
    \hspace*{-1mm}$t_{\mathrm{min}}$ & Minimum pulse width & $0.1$ $\mu$s \cite{skolnik2008radar} \\
\hline
\hspace*{-1mm}$\psi_l$ & Beamwidth of the ideal beam pattern & $\frac{\pi}{6}$ \cite{skolnik2008radar} \\
\hline
    \hspace*{-1mm}$\delta$ & Convergence tolerance & $10^{-3}$ \\
\hline
    \hspace*{-1mm}$\mu$ & Penalty factor & $10^{3}$ \\
\hline
\end{tabular}
\end{table}
\begin{figure}[t] 
\centering\includegraphics[width=3.2in]{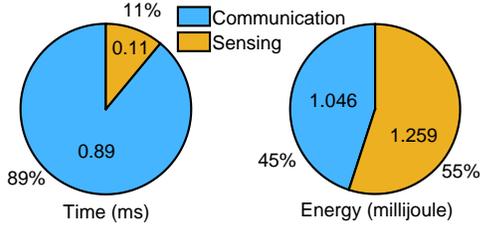}\vspace*{-8mm}
\caption{Illustration of the time and energy consumed for communication and sensing for an ISAC frame duration of $T=1$ ms.}\label{fig:Distri_Ant_ISAC_Pie}
\end{figure}
This section provides simulation results to evaluate the performance of the proposed DAN-based ISAC design. In particular, we assume that there are $J=3$ RRHs connected to the CP in the considered cell, as shown in Fig. \ref{fig:system_simulation}, and each RRH is equipped with $N_{\mathrm{T}}=6$ antennas. The three RRHs form an equilateral triangle with a side length of $500$ m while the CP is located at the centroid of the triangle. $K=3$ users and $L=3$ sensing locations are uniformly and randomly distributed in a disc with a radius of $500$ m centered at the location of the CP. The path loss exponents for the user channels and the channel between the RRHs and sensing locations are set to $3$ and $2$, respectively. The path loss at the reference distance of $1$ m is set to $40$ dB. We model the small-scale fading coefficients of the user channels as independent and identically distributed Rayleigh random variables. Each target location is assumed to be sensed by the nearest RRH employing a highly-directional beam. The angular domain is equally divided into $I=360$ directions to generate a set of $\{P_l(\theta_i)\}_{i=1}^{I}$ based on the pre-determined RRH-target pairs. To facilitate the presentation, we define the normalized beam pattern mismatch tolerance error $\overline{\epsilon}_l=\frac{\epsilon_l}{\sum_{i=1}^{I} P_l(\theta_i)}$ and set $\overline{\epsilon}_l=0.1$. The parameters adopted in the simulations are summarized in Table \ref{ISAC_parameters}. For comparison, we also consider two baseline schemes. For baseline scheme 1, we consider an ISAC system with co-located transmit antennas, where a conventional BS with $N_{\mathrm{T}}J$ antennas and a power budget $\sum_{j=1}^{J}P^{\mathrm{max}}_{\mathrm{T}_j}$ is located at the center of the cell. In this case, there is no need for fronthaul links and a multi-beam pattern is designed to concurrently sense all target locations. The total energy consumption of the considered system is minimized subject to constraints $\mbox{C1-C7}$ and $\mbox{C10-C12}$ of \eqref{prob1} by employing the proposed algorithm. For baseline scheme 2, we equally allocate the time to communication and sensing and solve \eqref{prob1} for $\eta=0.5$. The performance of the two baseline schemes will be shown in Fig. \ref{fig:Distri_Ant_ISAC_Energy}.
\subsection{Resource Allocation Policy of an ISAC Frame}
\par
\begin{figure}[t]
\centering\includegraphics[width=3.4in]{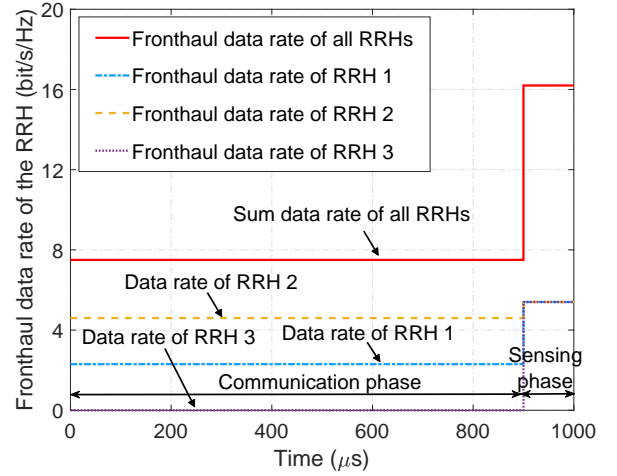} 
\caption{Illustration of the fronthaul data rate of each RRH and the sum data rate of all fronthaul links during the communication and sensing phases of the proposed scheme.} \label{fig:Distri_Ant_ISAC_Time}
\end{figure}

We first study the performance of the proposed scheme by focusing on a randomly generated channel realization. In Fig. \ref{fig:Distri_Ant_ISAC_Pie}, we show the time and energy used for communication and sensing for the proposed scheme with an ISAC frame duration of $T=1$ ms. We observe that most of the time (roughly $90 \%$) is assigned to the communication phase as this is favorable to satisfy the achievable rate requirement of the users in a power-efficient manner. On the other hand, although the time allocated to target sensing is much less than that for information transmission, the RRHs in the sensing phase consume more energy compared to the communication phase. This is because in the sensing phase, a comparatively high power is needed to combat the severe round-trip path loss such that the received power of the echoes at the RRHs exceeds the noise floor for reliable detection. Moreover, in Fig. \ref{fig:Distri_Ant_ISAC_Time}, we show the fronthaul data rate of each RRH and the sum data rate of all fronthaul links. We observe that compared to the communication phase, there is more data traffic between the RRHs and the CP in the sensing phase. This is because to acquire the sensing information of the target location in a power-efficient manner, the minimum sensing range has to be short to save power while the maximum sensing range should be larger than the distance of each RRH-target pair, leading to a large amount of quantized sensing data, cf. \eqref{sen_front}. Furthermore, we observe that during the communication phase, RRH $2$ serves two users in its vicinity while RRH $3$ is not used for information transmission as this is beneficial to improve power efficiency. In contrast, during the sensing phase, due to the pre-determined RRH-target pairs, each RRH senses a dedicated location and exchanges a similar amount of sensing data with the CP via the fronthaul link. 
\par
\begin{figure}[t] 
\centering\includegraphics[width=3.4in]{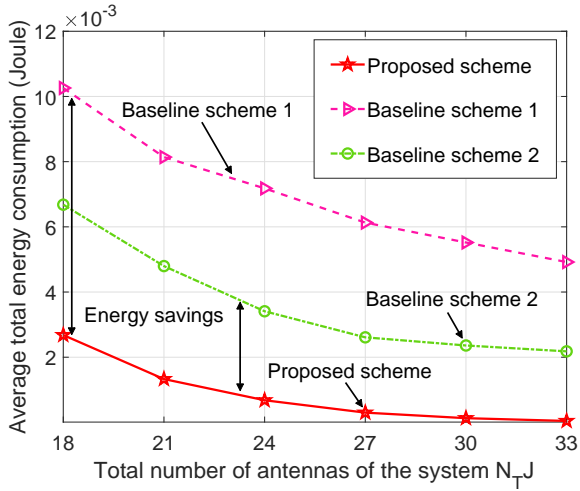}
\caption{Average total energy consumption during an ISAC frame (dBm) versus the total number of transmit antennas in the system, $N_{\mathrm{T}}J$, for different schemes.} \label{fig:Distri_Ant_ISAC_Energy}
\end{figure}

Fig. \ref{fig:Distri_Ant_ISAC_Energy} shows the total energy consumption in an ISAC frame (Joule) versus the total number of transmit antennas during the system for different schemes averaging over $200$ channel realization. As expected, for the proposed scheme and the two baseline schemes, the average total energy consumption during an ISAC frame decreases with $N_{\mathrm{T}}J$. In fact, by employing a larger set of antenna elements, the RRHs can exploit additional degrees-of-freedom (DoFs) to make the information beamforming power-efficient and to accurately synthesize the required energy-focused beam patterns, resulting in power savings. Moreover, compared to the proposed scheme, the two baseline schemes cause a significantly higher total energy consumption. This confirms the effectiveness of joint time, beamforming, and sensing signal optimization for the proposed DAN-based ISAC system. In particular, the co-located antenna architecture employed in baseline scheme 1 does not offer spatial macro-diversity to combat the path loss of the information and sensing signals. As for baseline scheme 2, although the RRHs can enjoy the performance gain introduced by the distributed antenna architecture, it is limited by the fixed duration of the communication and sensing phases.
\section{Conclusions}
In this paper, we studied the resource allocation algorithm design for DAN-based ISAC systems. In particular, to mitigate the interference between sensing and communication, we proposed to perform sensing and communication in two different phases. Taking into account the limited capacity of the fronthaul links and the QoS requirements of ISAC, we minimized the total energy consumption of all RRHs over a given time horizon by jointly optimizing the durations of both phases, information beamforming, and sensing signal. An computationally-efficient AO-based algorithm was developed for tackling the resulting non-convex optimization problem which allowed us to obtain a suboptimal solution. Simulation results showed that the proposed scheme can achieve significant energy savings compared to two baseline schemes. Moreover, our results revealed that compared to conventional networks with co-located antennas, the proposed DANs are a more promising architecture to facilitate future green ISAC systems.
\bibliographystyle{IEEEtran}
\bibliography{Reference_List}
\end{document}